\documentclass[conference]{IEEEtran}
\IEEEoverridecommandlockouts
\usepackage{cite}
\usepackage{amsmath,amssymb,amsfonts}
\usepackage{bm}
\usepackage{algorithmic}
\usepackage[linesnumbered,ruled,vlined]{algorithm2e}
\usepackage{graphicx}
\usepackage{textcomp}
\usepackage{xcolor}
\usepackage{capt-of}  
\usepackage{cuted}    
\usepackage{multirow}
\usepackage{lipsum}
\usepackage{hyperref}
\usepackage{caption}
\usepackage{subcaption}
\def\BibTeX{{\rm B\kern-.05em{\sc i\kern-.025em b}\kern-.08em
    T\kern-.1667em\lower.7ex\hbox{E}\kern-.125emX}}

\begin{document}

\title{Spectral Temporal Graph Neural Network for massive MIMO CSI Prediction\\}

\author{\IEEEauthorblockN{Sharan Mourya,} 
\and
\IEEEauthorblockN{Pavan Reddy,}
\and
\IEEEauthorblockN{SaiDhiraj Amuru,}
\and
\IEEEauthorblockN{Kiran Kumar Kuchi}
}


\maketitle

\begin{abstract}
In the realm of 5G communication systems, the accuracy of Channel State Information (CSI) prediction is vital for optimizing performance. This letter introduces a pioneering approach: the Spectral-Temporal Graph Neural Network (STEM GNN), which fuses spatial relationships and temporal dynamics of the wireless channel using the Graph Fourier Transform. We compare the STEM GNN approach with conventional Recurrent Neural Network (RNN) and Long Short-Term Memory (LSTM) models for CSI prediction. Our findings reveal a significant enhancement in overall communication system performance through STEM GNNs. For instance, in one scenario, STEM GNN achieves a sum rate of 5.009 bps/Hz which is $11.9\%$ higher than that of LSTM and $35\%$ higher than that of RNN. The spectral-temporal analysis capabilities of STEM GNNs capture intricate patterns often overlooked by traditional models, offering improvements in beamforming, interference mitigation, and ultra-reliable low-latency communication (URLLC). 
\end{abstract}


\section{Introduction}

In time division duplex (TDD) cellular networks, channel reciprocity enables downlink channel state information (DL-CSI) inference from uplink channel state information (UL-CSI), obviating the need for extra estimation or feedback. However, this reciprocity principle doesn't hold for frequency division duplex (FDD) systems, introducing challenges in DL-CSI acquisition. Prior methods required receivers to predict and feedback DL-CSI, incurring computational load and overheads. Addressing this, researchers have delved into channel prediction techniques. These techniques leverage historical CSI correlations to forecast future channel conditions, aiding multi-frame predictions. Existing methods mainly fall into model-based or neural network-based categories. Model-based strategies employ diverse models, like linear extrapolation, sum-of-sinusoids, and autoregressive (AR) \cite{csip4}, to capture dynamic channel behavior. However, complexities arising from factors like multipath and Doppler effects pose challenges to precise modeling. In contrast, neural network-driven approaches have gained traction \cite{csip4} which train neural networks to mirror real-world channel behaviors through data-driven learning.

In \cite{csip1}, authors employed feed-forward neural networks for CSI prediction, directly linking uplink and downlink CSI. However, their assumption of a bijective user-channel matrix relationship isn't always applicable. To address this, a more advanced model using Convolutional Neural Networks (CNN) and Long Short-Term Memory (LSTM) was proposed in \cite{csip2}, notably outperforming methods like Maximum Likelihood (ML) and Minimum Mean Squared Error (MMSE). A similar data-driven model based on CNN and LSTM was also presented by Wang \textit{et al.} \cite{csip3} for precise CSI prediction. Recurrent Neural Networks (RNNs) were employed in some strategies \cite{csip4} \cite{rnn}, often relying on CNNs for feature extraction, despite being primarily designed for real-valued image tasks. To alleviate this limitation, Zhang \textit{et al.} introduced a complex-valued 3D CNN for CSI prediction in \cite{csip5}, improving accuracy over real-valued networks. While LSTMs excel in capturing long-range correlations compared to RNNs, they face the vanishing gradient problem. Addressing this, a transformer-based CSI prediction model was introduced in \cite{csip6}. Transformers employ self-attention mechanisms, overcoming the vanishing gradient issue and preserving long-range correlations more effectively than LSTMs. Additionally, transformers resolve the sequential prediction problem, where predicting each future CSI matrix relies on historical CSI and the latest predicted CSI, leading to error propagation. Transformers process data in parallel, sidestepping the sequential prediction challenge and efficiently forecasting future CSI matrices.

Another category of deep learning architectures with high retention of long-range correlation is the Graph Neural Network (GNN) \cite{survey}. When dealing with long-range correlations, traditional neural networks like feed-forward or recurrent neural networks may struggle due to their local receptive fields and sequential nature. However, GNNs are well-suited to address this challenge as they incorporate message passing. GNNs operate by passing messages between nodes in a graph iteratively. Each node aggregates information from its neighboring nodes and updates its representation based on the gathered information. This process allows information to propagate across the graph, enabling nodes to capture correlations with more distant nodes. On the other hand, one of the fundamental components of GNNs is the graph convolutional layer \cite{survey}. This layer incorporates both local and non-local information, enabling nodes to capture correlations with their neighbors as well as more distant nodes in the graph. 


In this letter, we propose to use a spectral temporal (STEM) graph neural network \cite{main} for CSI prediction. Originally STEM GNN was proposed to solve the problem of time series forecasting \cite{survey1} on data with complex relational structures represented as graphs. STEM GNNs are particularly useful when dealing with multivariate time series data, where each time step contains multiple variables (e.g., temperature, humidity, pressure, etc.) and the interactions between variables can be represented as a graph. Our contributions in this letter can be summarized as follows:
\begin{enumerate}
    \item We model CSI prediction as a multivariate time-series forecasting problem, a popular class of problems in deep learning. 
    \item We modify the state-of-the-art STEM GNN model designed for multivariate time-series forecasting to better suit the CSI prediction problem. We exploit spectral and temporal correlations of the historical CSI with the future CSI to make accurate predictions.
    \item We implement other cutting-edge CSI prediction models using LSTMs and RNNs and make a detailed comparison with the STEM GNN approach.
\end{enumerate}
To the best of our knowledge, we are the first to model CSI prediction as a multivariate time series forecasting problem.
\begin{figure*}[h]
\centering
\includegraphics[width=0.8\linewidth]{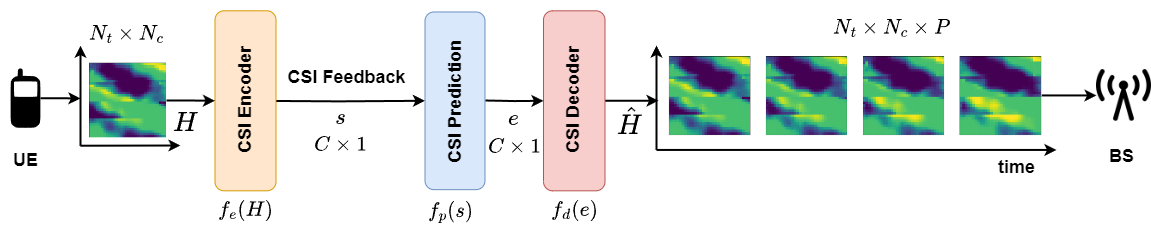}
\captionof{figure}{System Model} 
\label{model}
\end{figure*}

The organization of this paper is as follows: In Section II, we introduce the system model used throughout this paper. In Section III, we introduce STEM GNNs and their applicability to CSI prediction. In Section IV, we summarize the simulation results followed by the conclusion in Section V. 

\section{System Model}
We consider a MIMO orthogonal frequency division multiplexing (OFDM) system. In Fig.{~\ref{model}}, the base station (BS) is outfitted with an array of antennas organized in an $M\times N$ rectangular pattern, with each antenna featuring dual polarization resulting in a total of $N_{t} = M\times N\times 2$ transmit antennas and the user equipment (UE) has a single antenna. The number of sub-carriers is $N_{c}$. Given that the CSI prediction operates independently for various users, it is sufficient to illustrate the principle using a single user. In this context, we consider the communication channel between the user and the BS to be represented by a three-dimensional (3D) channel model, conforming to the specifications outlined in the 3rd Generation Partnership Project (3GPP) for 5G communications. This channel is denoted as ${H}$.

The received signal $y_{n}$ on the $n^{th}$ sub-carrier can be denoted by
\begin{equation}
    y_{n} = \textbf{${{h}_{n}}^{H}$}x_{n} + \textbf{$w_{n}$},
\end{equation}
where, ${h}_{n} \in \mathbb{C}^{N_{t}\times 1}$, $x_{n} \in \mathbb{C}$, and $w_{n} \in \mathbb{C}$ are the channel, symbol and additive noise respectively. The channel matrix ${H}$ of size $N_{c} \times N_{t}$ is given by
\begin{equation}
    {H} = \big[{h}_{1},{h}_{2},\cdots,{h}_{{N_{c}}} \big]^{H}.
\end{equation}
The matrix representing this channel consists of a total of $2N_{t}N_{C}$ elements, considering that the entries within the matrix are of a complex nature. However, in practical scenarios, transmitting such a sizable matrix proves to be unfeasible for a massive MIMO system. To address this challenge, we adopt a strategy to reduce the payload size. This entails compressing the matrix ${H}$, which is of a 3D nature, into a more compact one-dimensional vector with dimensions $C \times 1$, as depicted in Figure \ref{model}. To quantify this compression, we introduce the compression ratio denoted by $\gamma = \frac{C}{2N_{C}N_{t}}$. The compressed one-dimensional vector is then transmitted to the BS. Upon reception, a CSI prediction model is applied to this vector to forecast future CSI in its compressed form. Subsequently, a decoder is employed to reconstruct the estimated channel matrix, denoted as ${\hat{H}}$, from the predicted vector. This entire sequence of operations can be described through the following equations:
\[s = f_{e}({H}),\quad e = f_{p}(s),\quad {\hat{H}} = f_{d}(e),\]
where $f_{e}$ is the encoder, $f_{p}$ is the CSI prediction model, and $f_{d}$ is the decoder. $s$, $e$, and ${\hat{H}}$ are the compressed, predicted, and estimated channel matrices respectively.
Using the estimated channel matrix ${\hat{H}}$, BS utilizes the CSI prediction model to estimate the channel matrices for the future frames. We address the number of such frames as the horizon (denoted by $P$) and it would be specified in the CSI prediction model while training.

\section{Spectral Temporal Graph Neural Network}
Having established the system model, let's delve into the CSI prediction model depicted in Figure \ref{model}. Conventional CSI prediction relied on auto-regression \cite{ar} \cite{csip4} or hidden Markov chains \cite{markov}, which capture temporal correlations between historical and future CSI but neglect spatial matrix correlations. Thus, deep learning emerged, employing CNNs to capture the matrix's spatial relationships. Here, CSI is viewed as a spatial map, with each element representing a channel aspect. CNN-based models transform raw CSI into images highlighting patterns like signal fluctuations and interference.

Typically, CNN-based CSI models include an RNN or LSTM to capture temporal trends \cite{csip2, csip3, csip4, rnn}. Historical CSI sequences are fed into the network, learning patterns, and trends. RNNs and LSTMs retain memory, accurately inferring future states. However, this separate spatial-temporal approach misses the concurrent spatio-temporal correlations present in CSI evolution. We propose a unified model using GNNs, effectively extracting these correlations for efficient CSI prediction. STEM GNNs \cite{main} offer a profound approach to comprehending and predicting multivariate time-series patterns intertwined in space and time. Predicting such a series is intricate \cite{survey1}, requiring simultaneous consideration of intra-series and inter-series correlations. In CSI prediction, these map to spatial and temporal correlations. STEM GNNs adeptly handle both correlations simultaneously within the spectral domain. This is achieved through merging Graph Fourier Transform (GFT) \cite{survey} for spatial correlations and Discrete Fourier Transform (DFT) for temporal dependencies. These elements unite into an end-to-end framework, enabling STEM GNNs to effectively tackle both spatial and temporal correlations in CSI prediction.

\subsection{Problem Formulation}
\begin{figure}[h]
\centering
\includegraphics[width=0.7\linewidth]{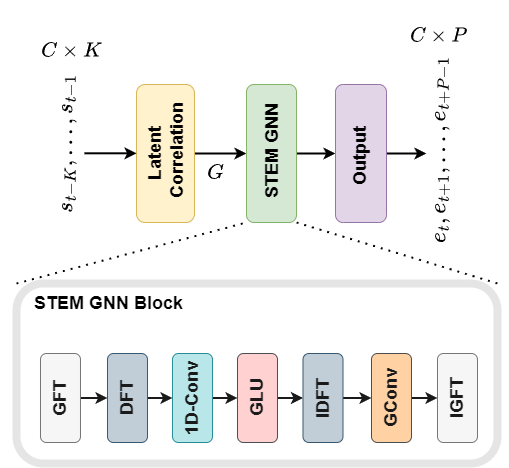}
\captionof{figure}{STEM GNN Architecture} 
\label{stemgnn}
\end{figure}
Let us establish the graph that captures the evolution of the channel as $G = (S, W)$, where $\{S = {s_{0}, s_{1}, \ldots, s_{T-1}}\} \in \mathbb{R}^{C\times T}$ signifies the historical compressed CSI input. In this context, $s(i,j)$ represents the latent representation vector, $C$ denotes its dimension (nodes), and $T$ represents the count of timestamps. The compressed CSI values observed at timestamp $t$ are symbolized as $s_{t} \in \mathbb{R}^{C}$. The adjacency matrix is denoted as $W \in \mathbb{R}^{C\times C}$, with $w_{ij} > 0$ indicating the presence of a connection between nodes $i$ and $j$, where the magnitude of $w_{ij}$ quantifies the strength of this connection.

Given the observed values from the $K$ preceding timestamps $\{{s_{t-K}, \ldots, s_{t-1}}\}$, the objective of CSI prediction revolves around anticipating the node values within a multivariate temporal graph $G = (S, W)$ for the subsequent $P$ timestamps, noted as $\{{e_{t}, e_{t+1}, \ldots, e_{t+P-1}}\}$. These values are deducible through the predictive model $f_{p}$ governed by parameters $\theta$.
\begin{equation}
    e_{t}, e_{t+1}, \hdots, e_{t+P-1} = f_{p}(s_{t-K}, \hdots, s_{t-1};G;\theta).
    \label{maineq}
\end{equation}
Now, using the decoder $f_{d}$ we decompress the predicted compressed CSI to obtain the future CSI as shown in Fig.{~\ref{model}}.

\subsection{STEM GNN Architecture}

STEM GNN consists of three main layers each having its own importance: latent correlation layer, STEM GNN layer, and output layer as shown in Fig.{~\ref{stemgnn}}. The latent correlation layer is tasked with finding the adjacency matrix ($W$) of the graph $G$. As it is difficult to construct an adjacency matrix of historical compressed CSI ($S$) by hand, a self-attention layer \cite{att} is used to learn the latent correlations between multiple compressed CSI vectors automatically. In this way, the model exploits correlations to form an adjacency matrix in a data-driven fashion. The extracted adjacency matrix $W$ along with the graph structure $G$ are fed to the next layer. STEM GNN layer has a couple of STEM GNN blocks (as shown in Fig.{~\ref{stemgnn}}) interconnected with skip connections and addition blocks \cite{main}. This forms the crust of the CSI prediction model as this takes $G$ as an input and extracts the spatio-temporal correlations from the historical compressed CSI which are then used by the output layer to create accurate predictions for the following $P$ timestamps.

The STEM GNN block encompasses several operations, including GFT and its inverse IGFT, DFT and its inverse IDFT, a 1D convolution layer (1D-Conv), followed by a gated linear unit (GLU) activation, and the pivotal graph convolution (GConv). In the initial steps, GFT transforms the graph $G$ into a spectral matrix representation, rendering compressed CSI independent in the spatial domain. Subsequently, DFT shifts univariate time-series components into the frequency domain, and 1D convolution with GLU captures feature patterns before reverting through inverse DFT. The cornerstone GConv operation, crucial in any GNN, propagates information across the graph. The sequence concludes with inverse GFT. This entire process is visually depicted in Figure \ref{stemgnn}.

\section{Results}
Utilizing the STEM GNN architecture and the problem formulation outlined earlier, we conducted a sequence of simulations using the 3GPP channel model, characterized by the following specifications.
\begin{table}[h]
\caption{Channel Model}
\centering
\begin{tabular}{|c|c|} 
 \hline
 Scenario & Urban Macro (UMa)\\
 \hline
 Antenna Config. & $[8, 2, 2]$ $\Rightarrow$ $N_{t} = 8\times 2\times 2 = 32$ \\ 
 \hline
 No. of Subcarriers ($N_{c}$) & 32\\
 \hline
 No. of consecutive timestamps ($T$) & 10,000\\
 \hline
 Bandwidth &20 MHz\\
 \hline
 User Mobility & 3Kmph\\
 \hline
\end{tabular}\\
\end{table}
The BS antenna is modelled by a uniform rectangular panel array, comprising $8 \times 2$ panels. The antenna panel is dual polarized hence the antenna configuration is $[8,2,2]$. We choose window size (K) in Eq.{~\ref{maineq}}, i.e., the number of previous timestamps to input to the CSI prediction model to be 12. Dimension of the compressed CSI, $C$ is chosen from $\{128,256,512\}$ which translates to compression ratio of $\gamma = \{1/16, 1/8, 1/4\}$ respectively. With these settings, we compare our approach with RNN and LSTM-based CSI prediction models as they outperform many of the traditional CSI prediction techniques \cite{csip4}. In addition, for CSI compression and feedback we use STNet \cite{self} as it has the best performance per resource utilization among all the other CSI feedback networks \cite{self_survey}.

\subsection{Sum Rate Performance}
Using transmit precoding, a massive MIMO system can attain remarkable capacities. In this regard, we employ the prevalent linear Zero-Forcing (ZF) transmit precoding technique \cite{self} to assess the overall enhancement in communication system performance brought about by distinct CSI feedback approaches. The evaluation process involves plotting the spectral efficiency of each method against the signal-to-noise ratio (SNR) for varying compression ratios. The resulting depiction can be observed in Figure \ref{stemgnn}.

\begin{figure}
     \centering
     \begin{subfigure}[b]{0.44\textwidth}
         \centering
         \includegraphics[width=\textwidth]{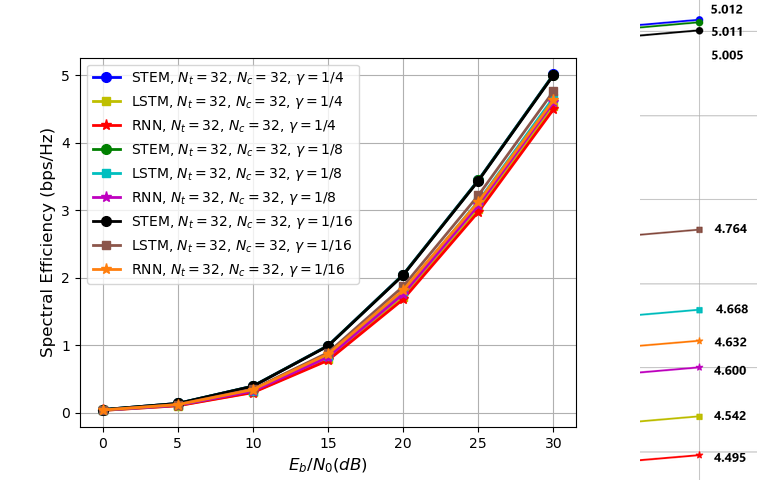}
         \caption{Horizon, $P = 5$}
         \label{se5}
     \end{subfigure}
     \hfill
     \begin{subfigure}[b]{0.44\textwidth}
         \centering
         \includegraphics[width=\textwidth]{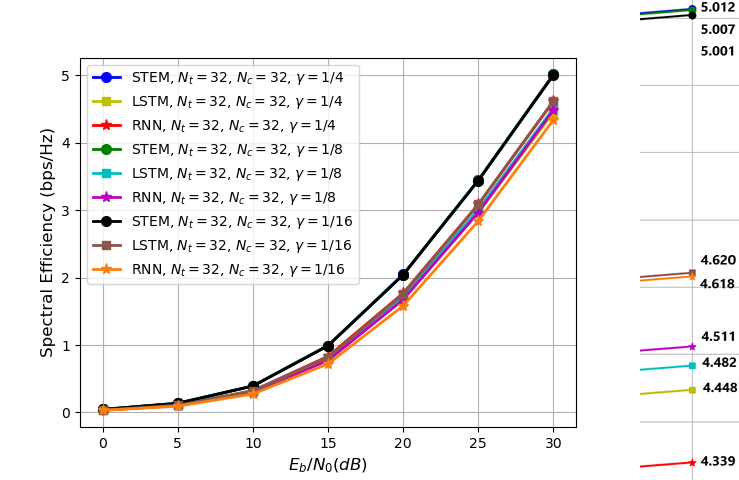}
         \caption{Horizon, $P = 7$}
         \label{se7}
     \end{subfigure}
    \hfill
     \begin{subfigure}[b]{0.44\textwidth}
         \centering
         \includegraphics[width=\textwidth]{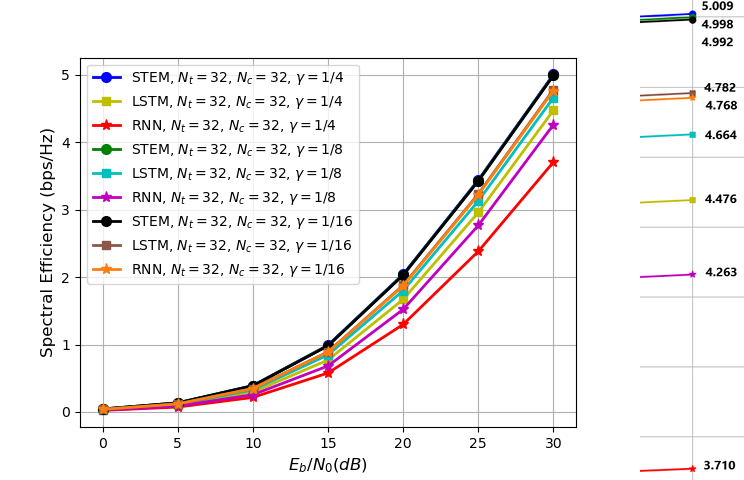}
         \caption{Horizon, $P = 9$}
         \label{se9}
     \end{subfigure}
        \caption{Sum rate vs SNR plots comparing STEM GNN with RNN and LSTM for CSI prediciton with various horizon values, $P = \{5,7,9\}$ at a fixed window size, $K=12$.}
        \label{se}
\end{figure}

Firstly, we note that for the STEM GNN approach, the sum rate increases with $\gamma$, and for RNN and LSTM, the sum rate decreases with $\gamma$. This is because RNNs and LSTMs lack the representational power of a GNN and hence fail to generalize the historical compressed CSI vectors for larger input sizes ($C$), i.e., smaller $\gamma$. On the other hand, STEM GNN is equipped with enough capabilities to generalize the compressed CSI for higher input sizes. Secondly, the difference in the sum rate performances of STEM GNN, RNN, and LSTM becomes wider and more evident as the horizon ($P$) value increases from $5$ to $9$ which is expected behaviour. For instance, at $P=9$, STEM GNN achieves a sum rate of 5.009 bps/Hz which is $11.9\%$ higher than that of LSTM and $35\%$ higher than that of RNN. Which, for a 20 MHz bandwidth translates to a total data rate of 74.2 MHz, 89.52 MHz, and 100.18 MHz for RNN, LSTM, and STEM GNN respectively. 

\subsection{CSI Prediction Performance}

\begin{figure}[h]
\centering
\includegraphics[width=\linewidth]{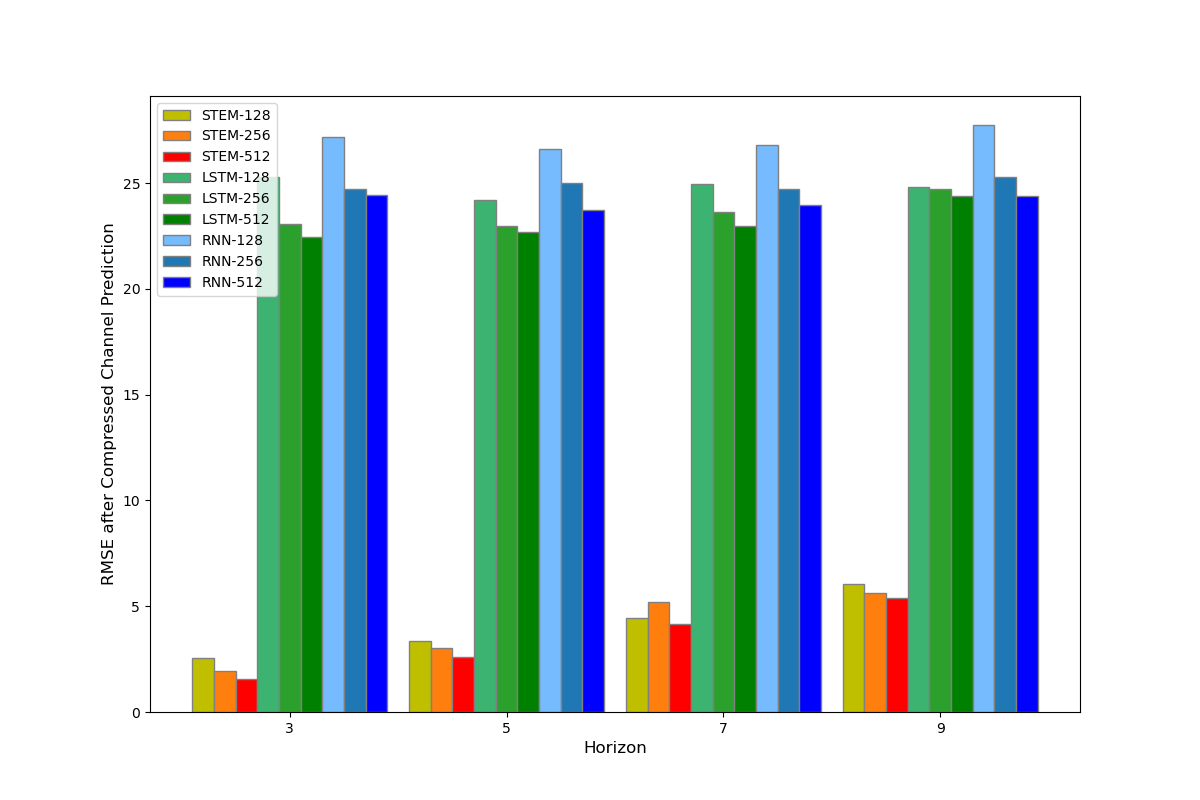}
\captionof{figure}{RMSE performance of RNN, LSTM, and STEM GNN-based CSI prediction models for various compression ratios ($\gamma$) and horizons ($P$).} 
\label{rmse}
\end{figure}

The accuracy of CSI prediction by various networks can be measured using root mean squared error (RMSE) \cite{main} between the original future CSI and the predicted future CSI. We plotted the RMSE values for RNN, LSTM, and STEM GNN networks for various compression ratios and horizons as shown in Fig {~\ref{rmse}}. It should be noted that the RMSE of STEM GNN is much lower when compared to RNN and LSTM networks. For instance, at $P=3$ and $\gamma=1/4$, STEM GNN achieved an RMSE of 1.5832 whereas RNN and LSTM achieved an RMSE of 24.4404 and 22.2726 which are 15 and 14-fold higher than that of STEM GNN which is a generational gap in performance.

\begin{figure}[h]
\centering
\includegraphics[width=\linewidth]{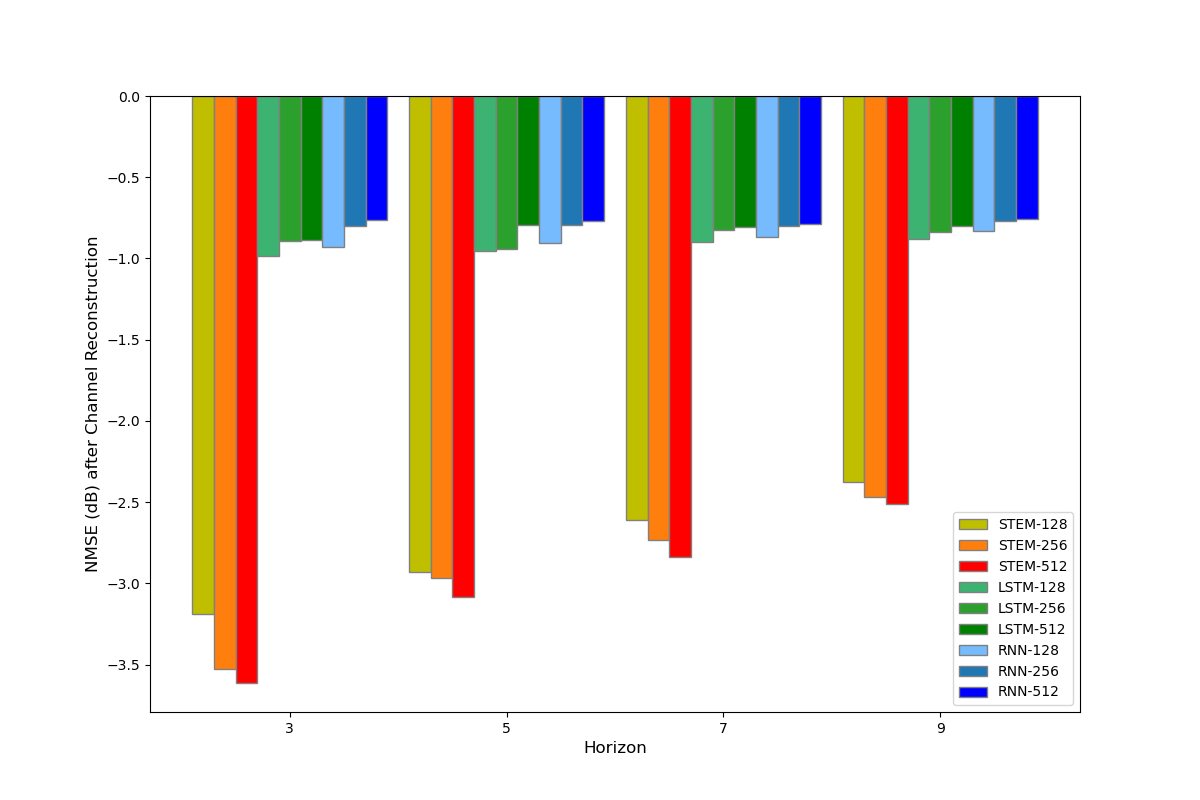}
\captionof{figure}{NMSE performance of RNN, LSTM, and STEM GNN-based CSI prediction models for various compression ratios ($\gamma$) and horizons ($P$).} 
\label{nmse}
\end{figure}

\begin{figure}[h]
\centering
\includegraphics[width=\linewidth]{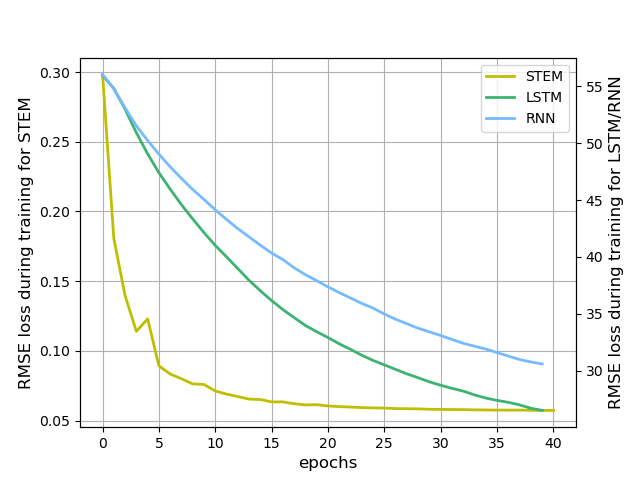}
\captionof{figure}{Training loss vs the number of epochs for RNN, LSTM, and STEM GNN-based CSI prediction models at the horizon, $P = 3$ and $\gamma = 1/4$.} 
\label{loss}
\end{figure}
\subsection{Channel Reconstruction Performance}
Similarly, the accuracy of CSI reconstruction at the BS can be measured using normalized mean squared error (NMSE) \cite{self} between the original CSI and the reconstructed CSI. This measure is important as its performance will inherently increase the system performance increasing the data rate. We plotted the NMSE values for RNN, LSTM, and STEM GNN networks for various compression ratios and horizons as shown in Fig.{~\ref{nmse}}. It should be noted that the NMSE of STEM GNN is much lower when compared to RNN and LSTM networks. For instance, at $P=3$ and $\gamma=1/4$, STEM GNN achieved an NMSE of -3.5116 whereas RNN and LSTM achieved an RMSE of -0.8967 and -0.7618 which are 3.9 and 4.6-fold higher than that of STEM GNN. This substantial difference in the NMSE performance is also reflected in the overall data rate achieved by different networks.

\subsection{Training Performance}
Training of all three networks was achieved on an Nvidia RTX 3060 graphic card\footnote{Source code of this paper: \url{https://github.com/sharanmourya/CSI-Prediction/}}. Training loss is an important parameter in understanding the stability and convergence of a network with respect to the dataset. Therefore, we plotted a training loss plot varying with the number of epochs for a horizon of $P=3$ and a compression ratio of $\gamma=1/4$ as shown in Fig.{~\ref{loss}}. From the figure, it can be seen that the training loss of STEM GNN is at least two orders of magnitude less than that of RNN and LSTM. In addition, the STEM GNN model also converges significantly faster than that of RNN and LSTM.

\section{Conclusion}
In this letter, we study the Spectral-Temporal Graph Neural Network (STEM GNN) approach for CSI prediction in 5G communication systems and demonstrate its superiority over traditional RNN and LSTM models. Leveraging both spatial relationships and temporal dynamics through the Graph Fourier Transform, STEM GNNs exhibit a remarkable improvement in overall communication system performance.

\bibliography{reference}
\bibliographystyle{ieeetr}
\end{document}